\documentclass{PoS} 
\usepackage{amsmath}\usepackage{amssymb}\usepackage{booktabs}
\title{OPE, Heavy-Quark Mass, and Heavy-Meson Decay Constants from
QCD Sum Rules}\ShortTitle{OPE, Heavy-Quark Mass, and Heavy-Meson
Decay Constants}\author{Wolfgang Lucha\\HEPHY, Austrian Academy of
Sciences, Nikolsdorfergasse 18, A-1050 Vienna, Austria\\E-mail:
\email{wolfgang.lucha@oeaw.ac.at}}\author{\speaker{Dmitri
Melikhov}\\HEPHY, Austrian Academy of Sciences, Nikolsdorfergasse
18, A-1050 Vienna, Austria\\Faculty of Physics, University of
Vienna, Boltzmanngasse 5, A-1090 Vienna, Austria\\SINP, Moscow
State University, 119991, Moscow, Russia\\E-mail:
\email{dmitri\_melikhov@gmx.de}}\author{Silvano Simula\\INFN,
Sezione di Roma III, Via della Vasca Navale 84, I-00146 Roma,
Italy\\E-mail: \email{simula@roma3.infn.it}}

\abstract{We present a sum-rule extraction of the decay constants
of heavy-light mesons from the two-point correlator of
pseudoscalar currents \cite{lms2010}. To this end, we compare the
perturbative expansions for the correlator and the decay constant
performed in terms of either the pole mass or the running
$\overline{\rm MS}$ mass of the heavy quark. The perturbative
expansion expressed in terms of the pole mass exhibits no sign of
convergence whereas reorganizing this very expansion in terms of
the $\overline{\rm MS}$ mass~yields a rather clear hierarchy of
the perturbative contributions. Accordingly, the decay constants
extracted from the pole-mass correlator turn out to be
considerably smaller than those extracted from its $\overline{\rm
MS}$-mass counterpart. Then, making use of the OPE in terms of the
$\overline{\rm MS}$ mass we derive the decay constants of heavy
mesons with emphasis on acquiring control over the uncertainties
in the decay constants, related both to the input QCD parameters
and to the limited accuracy of the method of sum rules. Gaining
this control has become possible due to the application of our
novel procedure for extracting hadron observables based on dual
thresholds which depend on the Borel parameter.}

\FullConference{The XIXth International Workshop on High Energy
Physics and Quantum Field Theory \\8--15 September
2010\\Golitsyno, Moscow, Russia}

\begin{document}

\section{Introduction: QCD Sum Rules, Quark--Hadron Duality, and
Effective Thresholds}The extraction of the decay constant of any
ground-state heavy pseudoscalar meson within the method of QCD sum
rules \cite{svz} poses, for the following well-known reasons, a
challenging problem:

First, one has to derive a reliable operator product expansion
(OPE) for the correlation function$$\Pi(p^2)\equiv{\rm i}\int{\rm
d}^4x\,\exp({\rm i}\,p\,x)\,
\langle0|T(j_5(x)\,j^\dagger_5(0))|0\rangle$$of two pseudoscalar
heavy-light currents $j_5(x)\equiv(m_Q+m)\,\bar q(x)\,{\rm
i}\,\gamma_5\,Q(x)$, with quark masses $m_Q$,~$m$.

Second, the knowledge of the truncated OPE for the correlator ---
even if the parameters of this OPE are known precisely --- allows
to extract the characteristics of the bound state with
only~limited accuracy which reflects the intrinsic uncertainties
of the method of QCD sum rules. Gaining control over these
uncertainties is a very subtle problem \cite{lms_1}.

Let us briefly recall the essential features of the sum-rule
computation of the decay constants. The quark--hadron duality
assumption yields a relation between ground-state contribution and
OPE for the Borel-transformed correlator with a cut applied at
some effective continuum threshold~$s_{\rm eff}$:\begin{equation}
\label{SR_QCD}f_Q^2\,M_Q^4\,\exp(-M_Q^2\,\tau)=\Pi_{\rm
dual}(\tau,s_{\rm eff})\equiv\hspace{-2.3ex}\int\limits^{s_{\rm
eff}}_{(m_Q+m)^2}\hspace{-2.3ex}{\rm
d}s\,\exp(-s\,\tau)\,\rho_{\rm pert}(s)+\Pi_{\rm
power}(\tau),\end{equation}where the perturbative spectral density
is obtained as expansion in powers of the strong
coupling~$\alpha_{\rm s}$:$$\rho_{\rm pert}(s)=\rho^{(0)}(s)
+\frac{\alpha_{\rm s}}{\pi}\,\rho^{(1)}(s)+
\left(\frac{\alpha_{\rm s}}{\pi}\right)^2\,\rho^{(2)}(s)+\cdots.$$
Evidently, in order to extract the decay constant one has to fix
the effective continuum threshold~$s_{\rm eff}$.

A crucial albeit rather obvious observation is that $s_{\rm eff}$
has to be a function of $\tau$. Otherwise the l.h.s.\ and the
r.h.s.\ of (\ref{SR_QCD}) exhibit a different $\tau$-behavior. The
{\em exact effective continuum threshold\/}, corresponding to
exact values of hadron mass and decay constant on the l.h.s. of
(\ref{SR_QCD}), is, of course, not known. Therefore, the
extraction of hadron parameters from the sum rule consists in
attempting (i) to find a reasonable approximation to the exact
threshold and (ii) to control the accuracy of this approximation.
In a series of publications \cite{lms_new} we have formulated all
the corresponding procedures.

Let us introduce the dual invariant mass $M_{\rm dual}$ and the
dual decay constant $f_{\rm dual}$ by the relations
\begin{equation}\label{mf-dual}M_{\rm dual}^2(\tau)\equiv
-\frac{{\rm d}}{{\rm d}\tau}\log\Pi_{\rm dual}(\tau,s_{\rm
eff}(\tau)), \qquad f_{\rm dual}^2(\tau)\equiv
M_Q^{-4}\,\exp(M_Q^2\,\tau)\,\Pi_{\rm dual}(\tau,s_{\rm
eff}(\tau)).\end{equation}If the ground-state mass is known, any
deviation of the dual mass from the actual ground-state mass
yields an indication of the amount of excited-state contribution
picked up by our dual correlator. Assuming a particular behavior
of the effective threshold with $\tau$ and requiring the least
deviation of the dual mass in (\ref{mf-dual}) from the actual mass
$M_Q$ in the $\tau$-window yields a variational solution for the
effective threshold $s_{\rm eff}(\tau)$. As soon as the latter has
been fixed we get the decay constant from (\ref{mf-dual}). The
standard assumption for the effective threshold is a
$\tau$-independent constant. In addition to this approximation, we
have considered polynomials in $\tau$. Reproducing the actual mass
considerably improves for $\tau$-dependent thresholds. This means
that a dual correlator with $\tau$-dependent threshold isolates
the ground-state contribution much better and is less contaminated
by the excited states than a dual correlator with standard
$\tau$-independent threshold. As consequence, the accuracy of
extracted hadron observables is significantly increased. Recent
experience from potential models reveals that the band of values
obtained from the linear, quadratic, and cubic Ans\"atze for the
effective threshold encompasses the true value of the decay
constant \cite{lms_new}. Moreover, we managed to demonstrate that
the extraction procedures in quantum mechanics and QCD are even
quantitatively rather similar \cite{lms_qcdvsqm}. This
contribution reports our recent findings
\cite{lms2010,lms2010_conf} for pseudoscalar-heavy-meson decay
constants.

\section{OPE and Heavy-Quark Mass}
\begin{figure}[!h]\begin{tabular}{cc}
\includegraphics[width=7cm]{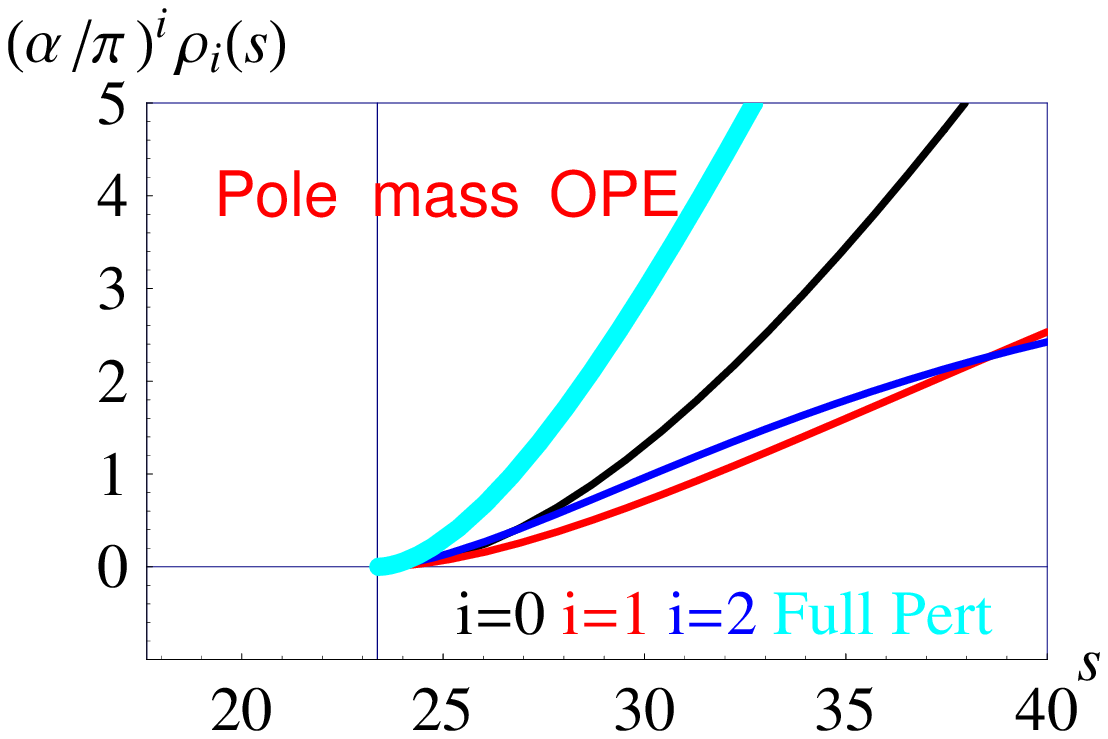}&
\includegraphics[width=7cm]{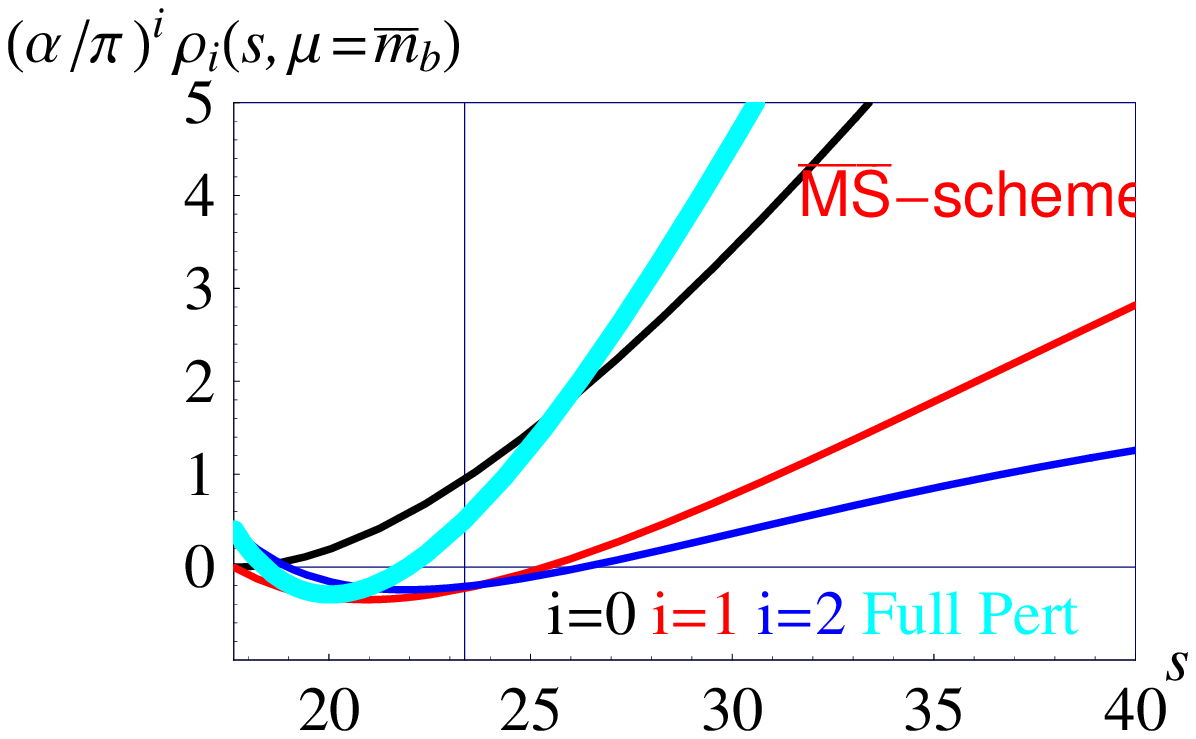}\\
\includegraphics[width=7cm]{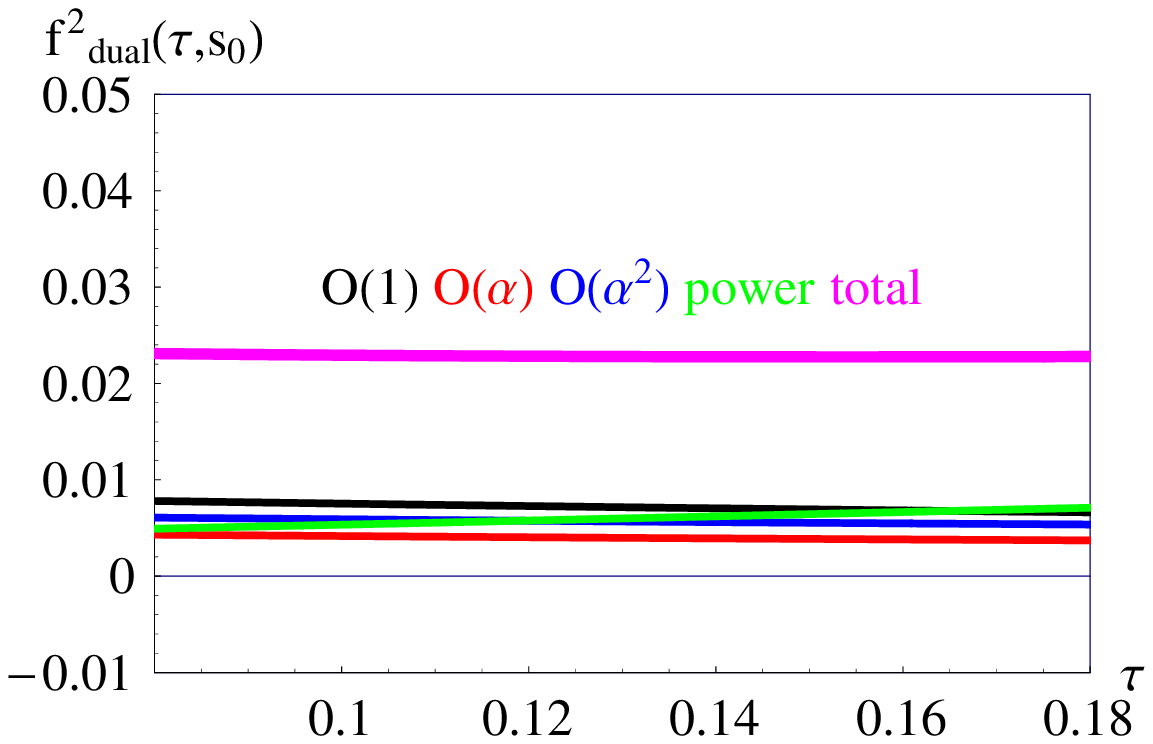}&
\includegraphics[width=7cm]{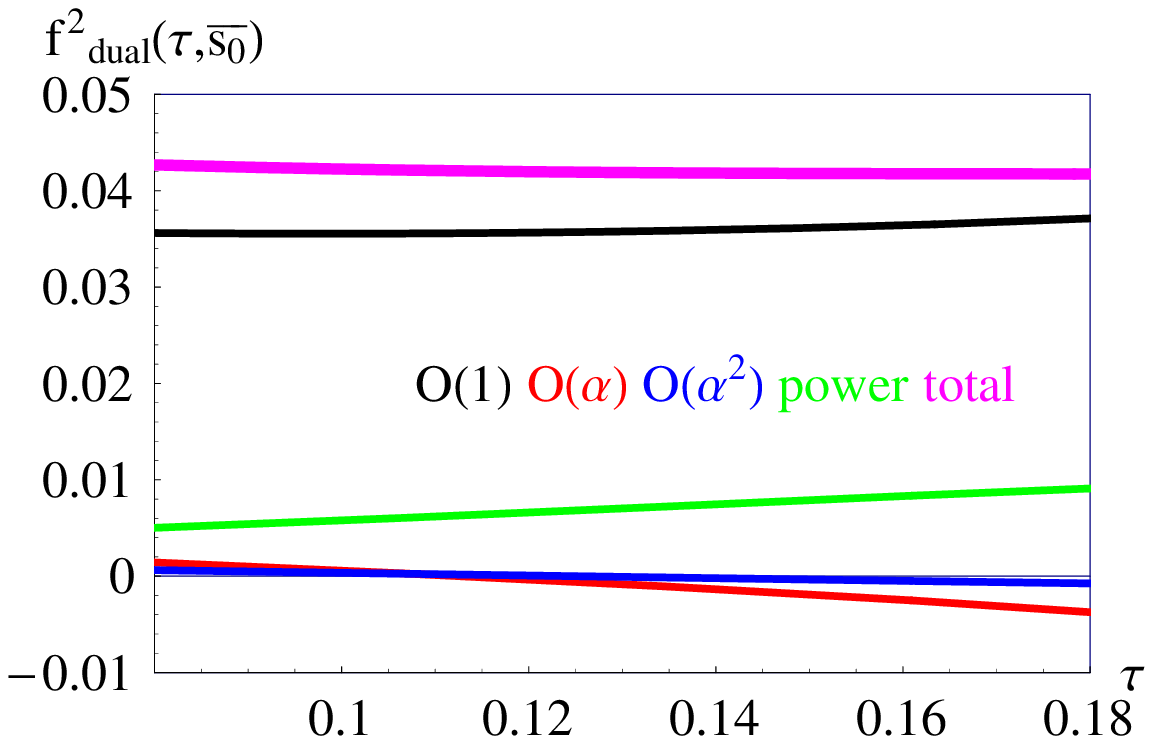}\end{tabular}
\caption{\label{Plot:1}OPE expressed in terms of the pole mass
(left) and the $\overline{\rm MS}$ mass (right) of the $b$ quark:
First row: spectral densities. Second row: corresponding sum-rule
estimates for $f_B$; a constant effective continuum threshold is
fixed in each case by requiring maximum stability of the extracted
decay constant, thus $s_0\ne \bar s_0$.}\end{figure}

For heavy-light correlators and
decay constants it makes a big difference which scheme for the
heavy-quark mass is used. We adopt the OPE for this correlator to
three-loop accuracy \cite{chetyrkin}, which was obtained in terms
of the pole mass of the heavy quark. The pole-mass scheme is the
standard one; it has been used for a long time since the
pioneering work of Ref.~\cite{aliev}. An alternative is to
reorganize the perturbative expansion in terms of the running
$\overline{\rm MS}$ mass \cite{jamin}. Since the correlator is
known to $\alpha_{\rm s}^2$ accuracy, the relation between the
pole and the $\overline{\rm MS}$ mass to the same accuracy is
used. Figure \ref{Plot:1} depicts the corresponding spectral
densities and our sum-rule $f_B$ estimates for both cases.

Several important lessons may be learnt from the instructive plots
in Fig.~\ref{Plot:1}:

\vspace{1.5ex}\noindent(i) The perturbative expansion for the
decay constant in terms of the pole mass exhibits no sign of
convergence; each of the terms --- LO, NLO, NNLO --- gives a
positive contribution of similar size. Thus, there is no reason to
expect that higher orders give smaller contributions. As a
consequence, the decay constant extracted from the pole-mass OPE
considerably underestimates the actual value.

\vspace{1.5ex}\noindent(ii) Reorganizing the perturbative series
in terms of the $\overline{\rm MS}$ mass of the heavy quark leads
to a clear hierarchy of the perturbative contributions
\cite{jamin}. Notice, however, that also in this case the
situation is not perfect: the full spectral density --- which is a
positive-definite function --- is negative at small values of $s$.
This is an artifact of the truncation and indicates that the
contributions of higher-order terms are non-negligible.

\vspace{1.5ex}\noindent(iii) The absolute value of the decay
constant obtained from the pole-mass OPE proves to be {\em almost
50\% smaller\/} than in the case of the $\overline{\rm MS}$
scheme. Let us emphasize that, nevertheless, in both cases the
decay constant exhibits perfect stability over a wide range of the
Borel parameter $\tau$. Thus, mere {\bf Borel stability is, by
far, not sufficient to guarantee the reliability of the sum-rule
extraction of bound-state parameters.} We have pointed out this
fact already more than once \cite{lms_1}. Unfortunately, some
authors still adhere to the idea that Borel stability is a
``proof'' of the reliability of their results.

\vspace{1.5ex}Because of the obvious problems with the pole-mass
OPE for the correlator, following \cite{jamin} we make use of the
OPE formulated in terms of the $\overline{\rm MS}$ mass for the
extraction of the decay constants.

\section{Decay Constants of $D$ and $D_s$ Mesons}
\vfill\begin{figure}[!b]\begin{center}\begin{tabular}{cc}
\includegraphics[width=7cm]{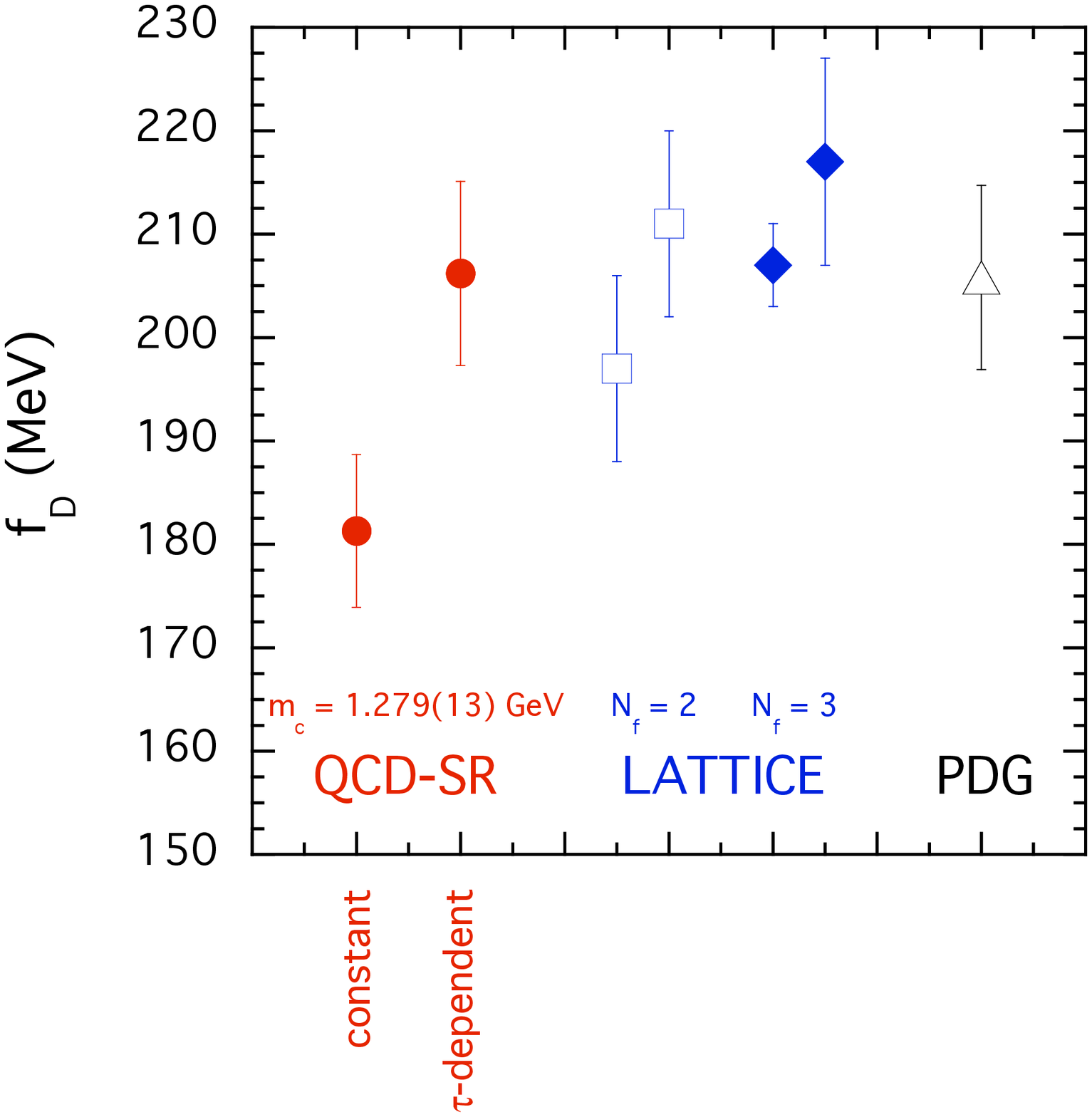}\hspace{1ex}&\hspace{1ex}
\includegraphics[width=7cm]{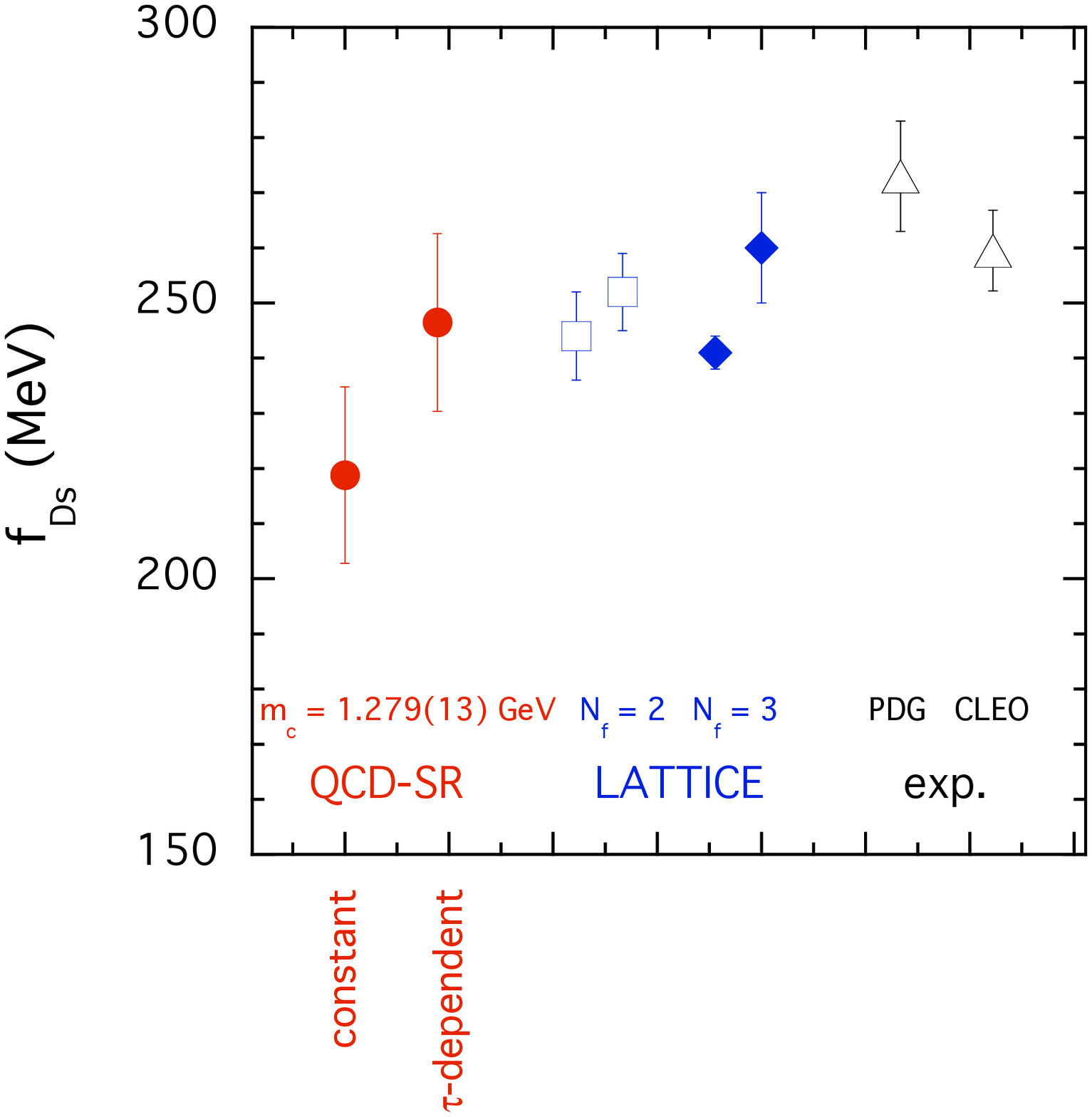}\end{tabular}\end{center}
\caption{\label{Plot:Dresults}Comparison of our findings for $f_D$
and $f_{D_s}$ with lattice results and experiment; for details,
see \cite{lms2010}.}\end{figure}

The application of our modified extraction prescriptions relying
on the $\tau$-dependent effective threshold leads to the following
results for charmed mesons (a detailed analysis can be found
in~\cite{lms2010}):\begin{align*}
f_{D}&=(206.2\pm7.3_{\rm(OPE)}\pm5.1_{\rm(syst)})\;\mbox{MeV},\\
f_{D_s}&=(245.3\pm15.7_{\rm(OPE)}\pm4.5_{\rm(syst)})\;{\rm MeV}.
\end{align*}The OPE-related error is obtained by bootstrapping,
allowing for a variation of all QCD parameters (i.e., quark
masses, $\alpha_{\rm s}$, condensates) in the relevant ranges. One
observes a perfect agreement~of~our predictions with the
respective lattice results (Fig.~\ref{Plot:Dresults}). It should
be emphasized that our $\tau$-dependent threshold is a crucial
ingredient for a successful extraction of the decay constants from
the sum rule. Obviously, the standard $\tau$-independent na\"ive
approximation yields a much lower value for $f_D$ which lies
rather far from the empirical data deduced from experiment {\em
and\/} from the lattice computations.

Let us emphasize the following: In quantum-mechanical potential
models \cite{lms_new} we succeeded to show that taking into
account the $\tau$-dependence of the effective threshold
considerably improves the accuracy of the duality approximation
and the quality of the sum-rule estimates. {\bf The investigation
of the decay constant of the $D$ meson clearly demonstrates that
also in QCD the $\tau$-dependent threshold leads to a much better
accuracy of the duality approximation (see Figs.~1, 3, 5, and 8 of
Ref.~\cite{lms2010}) and to a discernible improvement of the
accuracy of the extracted decay constant.} This perfectly confirms
our observation \cite{lms_qcdvsqm} that the extraction procedures
in QCD and in quantum mechanics are very similar to each other ---
both qualitatively and quantitatively. Moreover, our $D$ meson
analysis gives strong arguments that our algorithm provides quite
realistic systematic errors.

\section{$\overline{\rm MS}$ Mass of the $b$ Quark and Decay
Constants of $B$ and $B_s$ Mesons}
\begin{figure}[!b]\begin{center}\begin{tabular}{cc}
\includegraphics[width=6.387cm]{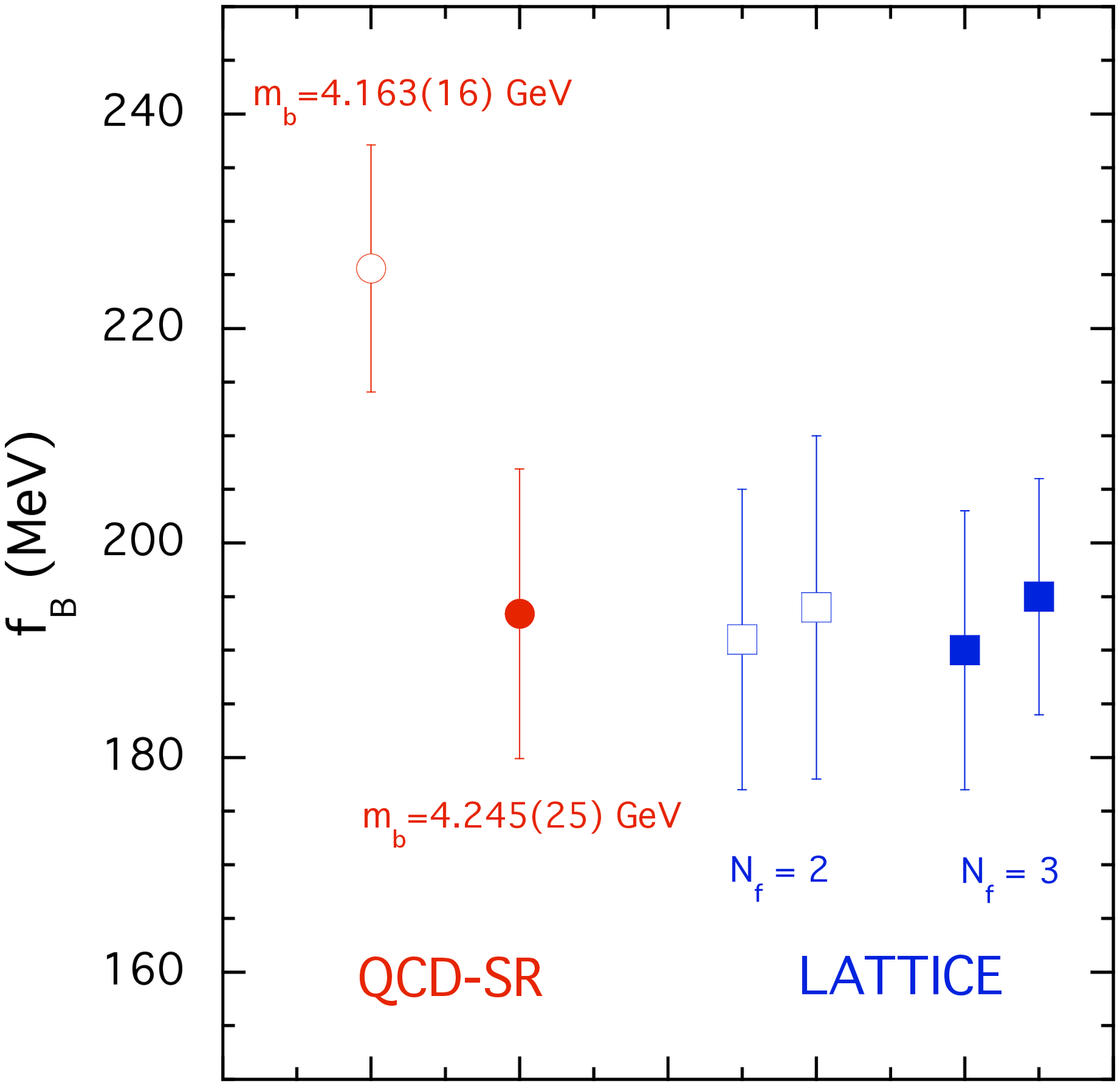}\hspace{1ex}&\hspace{1ex}
\includegraphics[width=6.387cm]{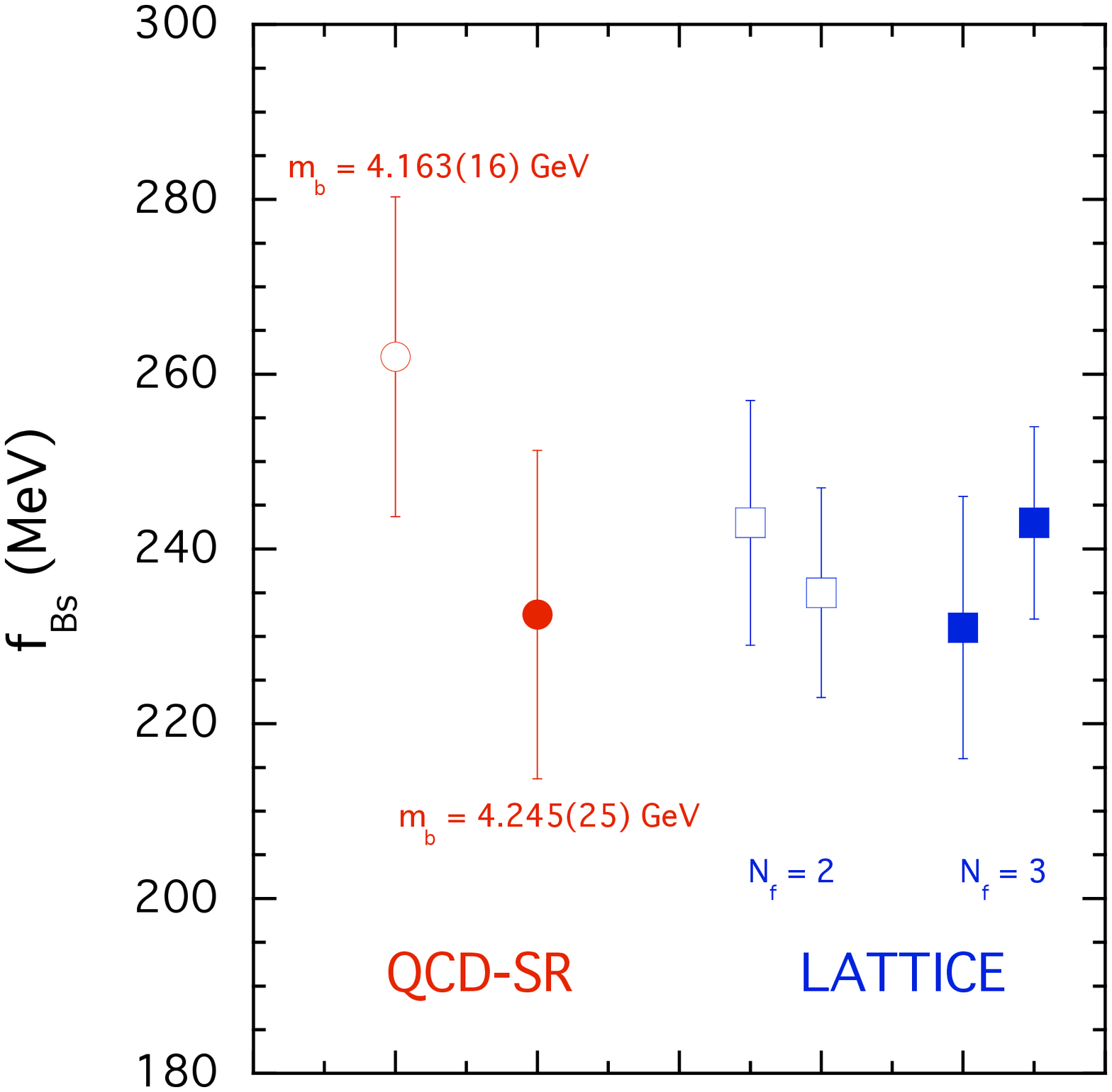}\end{tabular}\end{center}
\caption{\label{Plot:Bresults}Comparison of our results for $f_B$
and $f_{B_s}$ with lattice results; for a detailed list of
references,~cf.~\cite{lms2010}.}\end{figure}

The values of the beauty-meson
decay constants extracted from QCD sum rules are extremely
sensitive to the chosen $b$ mass $\overline{m}_b(\overline{m}_b)$.
For instance, the range $\overline{m}_b(\overline{m}_b)=(4.163\pm
0.016)\;\mbox{GeV}$ \cite{mb} yields results for the decay
constants that are barely compatible with the lattice
calculations~(Fig.~\ref{Plot:Bresults}). Requiring our sum-rule
result for $f_B$ to match the average of the lattice
determinations entails the rather precise value of the $b$-quark
mass$$\overline{m}_b(\overline{m}_b)=(4.245\pm0.025)\;\mbox{GeV}.$$
Our sum-rule estimates for $f_B$ and $f_{B_s}$ corresponding to
this value of the $b$-quark mass thus become\begin{align*}
f_{B}&=(193.4\pm12.3_{\rm(OPE)}\pm4.3_{\rm(syst)})\;\mbox{MeV},\\
f_{B_s}&=(232.5\pm18.6_{\rm(OPE)}\pm2.4_{\rm(syst)})\;\mbox{MeV}.
\end{align*}

\vfill

\section{Summary and Conclusions}The main insights gained in
our comprehensive inspection
\cite{lms2010,lms_1,lms_new,lms_qcdvsqm,lms2010_conf} of the QCD
sum-rule approach and its potential improvement by reasonable
modifications can be summarized as follows:

\vspace{1.5ex}\noindent1.~The $\tau$-dependence of the effective
continuum threshold emerges naturally when one attempts to make
the duality relation exact: this dependence is obvious from
(\ref{SR_QCD}). We would like to emphasize the following two
points: (a) In principle, the $\tau$-dependence cannot be in
conflict with any property of quantum field theories. (b) Many
examples, in particular, the analysis of the decay constants of
$D$ mesons presented here, show that it indeed improves visibly
the quality of the sum-rule predictions.

\vspace{1.5ex}\noindent2.~Our analysis of {\em charmed mesons\/}
clearly demonstrates that the use of Borel-parameter-dependent
thresholds leads to two essential improvements: (i) The accuracy
of decay constants extracted from sum rules is considerably
improved. (ii) It has become possible to obtain realistic
systematic errors and to reduce their values to the level of a few
percent. The application of our prescription brings the QCD
sum-rule results into perfect agreement with the findings of both
lattice QCD and experiment.

\vspace{1.5ex}\noindent3.~The {\em beauty-meson\/} decay constants
prove to be extremely sensitive to the chosen value of
$\overline{m}_b(\overline{m}_b)$; matching the results from QCD
sum rules for $f_B$ to the average of the lattice evaluations
allows us to provide a rather accurate estimate of the $b$-quark
mass. Our $\overline{m}_b(\overline{m}_b)$ value is in good
agreement with several lattice results but, interestingly, does
not overlap with the recent accurate determination presented in
Ref.~\cite{mb} (for details, see Ref.~[1]). Of course, this
intriguing puzzle should be clarified.

\vspace{3ex}\noindent{\bf Acknowledgments.} D.M.\ was supported by
the Austrian Science Fund (FWF), project no.~P20573.


\begin{thebibliography}{99}
\bibitem{lms2010}W.~Lucha, D.~Melikhov, and S.~Simula,
arXiv:1008.2698 [hep-ph].
\bibitem{svz}M.~Shifman, A.~Vainshtein, and V.~Zakharov,
Nucl.~Phys.~B {\bf 147}, 385 (1979).
\bibitem{lms_1}W.~Lucha, D.~Melikhov, and S.~Simula, Phys.~Rev.~D
{\bf 76}, 036002 (2007); Phys.~Lett.~B {\bf 657}, 148 (2007);
Phys.~Atom.~Nucl.~{\bf 71}, 1461 (2008); Phys.~Lett.~B {\bf 671},
445 (2009); D.~Melikhov, Phys.~Lett.~B {\bf 671}, 450 (2009).
\bibitem{lms_new}W.~Lucha, D.~Melikhov, and S.~Simula, Phys.~Rev.~D
{\bf 79}, 096011 (2009); J.~Phys.~G: Nucl.~Part.~Phys.\ {\bf 37},
035003 (2010); W.~Lucha, D.~Melikhov, H.~Sazdjian, and S.~Simula,
Phys.~Rev.~D {\bf 80}, 114028 (2009).
\bibitem{lms_qcdvsqm}W.~Lucha, D.~Melikhov, and S.~Simula,
Phys.~Lett.~B {\bf 687}, 48 (2010); Phys.~Atom.~Nucl.~{\bf 73},
1770 (2010).
\bibitem{lms2010_conf}W.~Lucha, D.~Melikhov, and S.~Simula,
arXiv:1008.2951 [hep-ph]; arXiv:1008.3129 [hep-ph];
arXiv:1011.3372 [hep-ph].
\bibitem{chetyrkin}K.~G.~Chetyrkin and M.~Steinhauser,
Phys.~Lett.~B {\bf 502}, 104 (2001); Eur.~Phys.~J.~C {\bf 21}, 319
(2001).
\bibitem{aliev}T.~M.~Aliev and V.~L.~Eletsky, Yad.~Fiz.~{\bf 38},
1537 (1983).
\bibitem{jamin}M.~Jamin and B.~O.~Lange, Phys.~Rev.~D {\bf 65},
056005 (2002).
\bibitem{mb}K.~G.~Chetyrkin {\em et al\/}., Phys.~Rev.~D {\bf 80},
074010 (2009).
\end{thebibliography}
\end{document}